\pdfoutput=1
\documentclass[useAMS,usenatbib,letters]{mn2e}
\usepackage[pdftex]{graphicx}
\usepackage[varg]{txfonts}
\usepackage{wasysym}
\usepackage{url}
\usepackage{lineno}
\usepackage{amssymb}
\usepackage{subfigure}

\newcommand{\etacar}{$\eta$~Car}
\newcommand{\g}{$\gamma$}

\newcommand{\lumi}{\,erg\,s$^{-1}$}

\newcommand{\fermi}{\emph{Fermi}-LAT}

\title{On the origin of $\gamma$-ray emission in $\eta$ Carina}

\author[S. Ohm, V. Zabalza, J.A. Hinton and E.R. Parkin]{\normalsize
  S.~Ohm,$^{1}$, V. Zabalza$^{2}$, J.A.~Hinton$^{2,3}$ and E.R. Parkin$^{4}$
  \\
  $^{1}$
  DESY, D-15738 Zeuthen, Germany \\
  $^{2}$
  Department of Physics and Astronomy, The University of Leicester, University Road, Leicester, LE1 7RH, United Kingdom \\
  $^{3}$
  Max Planck Institut f\"ur Kernphysik, Heidelberg D-69029, Germany \\
  $^{4}$
  School of Physics and Astronomy, University of Leeds, Woodhouse
  Lane, Leeds, LS2 9JT, United Kingdom\\
}

\begin{document}

\date{Accepted 2015 February 13. Received 2015 February 13; in original form
  2015 January 28}
\pagerange{\pageref{firstpage}--\pageref{lastpage}} \pubyear{2015}

\maketitle

\label{firstpage}

\begin{abstract}
  \etacar\ is the only colliding-wind binary for which high-energy \g\
  rays are detected. Although the physical conditions in the shock
  region change on timescales of hours to days, the variability seen
  at GeV energies is weak and on significantly longer timescales. The
  \g-ray spectrum exhibits two features that can be interpreted as
  emission from the shocks on either side of the contact
  discontinuity. Here we report on the first time-dependent modelling
  of the non-thermal emission in \etacar. We find that emission from
  primary electrons is likely not responsible for the \g-ray emission,
  but accelerated protons interacting with the dense wind material can
  explain the observations.  In our model, efficient acceleration is
  required at both shocks, with the primary side acting as a hadron
  calorimeter, whilst on the companion side acceleration is limited by
  the flow time out of the system, resulting in changing acceleration
  conditions. The system therefore represents a unique laboratory for
  the exploration of hadronic particle acceleration in
  non-relativistic shocks.
\end{abstract}


\maketitle

%
%
\def\aj{AJ}%
\def\actaa{Acta Astron.}%
\def\araa{ARA\&A}%
\def\apj{ApJ}%
\def\apjl{ApJ}%
\def\apjs{ApJS}%
\def\ao{Appl.~Opt.}%
\def\apss{Ap\&SS}%
\def\aap{A\&A}%
\def\aapr{A\&A~Rev.}%
\def\aaps{A\&AS}%
\def\azh{AZh}%
\def\baas{BAAS}%
\def\bac{Bull. astr. Inst. Czechosl.}%
\def\caa{Chinese Astron. Astrophys.}%
\def\cjaa{Chinese J. Astron. Astrophys.}%
\def\icarus{Icarus}%
\def\jcap{J. Cosmology Astropart. Phys.}%
\def\jrasc{JRASC}%
\def\mnras{MNRAS}%
\def\memras{MmRAS}%
\def\na{New A}%
\def\nar{New A Rev.}%
\def\pasa{PASA}%
\def\pra{Phys.~Rev.~A}%
\def\prb{Phys.~Rev.~B}%
\def\prc{Phys.~Rev.~C}%
\def\prd{Phys.~Rev.~D}%
\def\pre{Phys.~Rev.~E}%
\def\prl{Phys.~Rev.~Lett.}%
\def\pasp{PASP}%
\def\pasj{PASJ}%
\def\qjras{QJRAS}%
\def\rmxaa{Rev. Mexicana Astron. Astrofis.}%
\def\skytel{S\&T}%
\def\solphys{Sol.~Phys.}%
\def\sovast{Soviet~Ast.}%
\def\ssr{Space~Sci.~Rev.}%
\def\zap{ZAp}%
\def\nat{Nature}%
\def\iaucirc{IAU~Circ.}%
\def\aplett{Astrophys.~Lett.}%
\def\apspr{Astrophys.~Space~Phys.~Res.}%
\def\bain{Bull.~Astron.~Inst.~Netherlands}%
\def\fcp{Fund.~Cosmic~Phys.}%
\def\gca{Geochim.~Cosmochim.~Acta}%
\def\grl{Geophys.~Res.~Lett.}%
\def\jcp{J.~Chem.~Phys.}%
\def\jgr{J.~Geophys.~Res.}%
\def\jqsrt{J.~Quant.~Spec.~Radiat.~Transf.}%
\def\memsai{Mem.~Soc.~Astron.~Italiana}%
\def\nphysa{Nucl.~Phys.~A}%
\def\physrep{Phys.~Rep.}%
\def\physscr{Phys.~Scr}%
\def\planss{Planet.~Space~Sci.}%
\def\procspie{Proc.~SPIE}%

\section{Introduction}\label{sec:intro}

Particle acceleration up to very high energies in pulsar wind nebulae
and supernova remnants is well established, with non-thermal emission
seen from radio to TeV energies. The mechanism of diffusive shock
acceleration \citep[DSA; e.g.][]{Drury1983} is usually invoked to
explain the non-thermal emission in these systems, suggesting that
{all} systems with strong shocks accelerate particles. The shocks
present in some Galactic colliding wind binaries (CWBs) -- binary
systems with two massive stars and powerful winds -- seem to satisfy
all the criteria for particle acceleration and detectable high energy
emission: shock velocities of $\gtrsim$1000\,km\,s$^{-1}$, available
wind power of $\sim$$10^{37}$\,erg\,s$^{-1}$,
and copious targets for the production of high-energy radiation: soft
photon fields and high-density gas. Indeed non-thermal radio emission
from some CWBs has been seen \citep[e.g.][]{CWB:Review}, and models
predicting emission at \g-ray energies from such systems developed
\citep[e.g.][]{Eichler1993, Benaglia03, CWB:Bednarek05,
  CWB:Reimer06}. Only recently, however, has strong experimental
evidence appeared for \g-ray emission from such systems: in the unique
case of \etacar, using AGILE \citep{EtaCar:Agile} and \fermi\
\citep[][]{Fermi:BSL}. A hard X-ray tail is also seen from \etacar\
\citep{Viotti2004, EtaCar:Leyder10}.

\etacar\ is a binary system in a $\sim$5.5-year
orbit \citep{Damineli2008}. The masses and mass-loss rates of the
stellar companions (Table~\ref{tab:stars}), together with the
eccentricity of the CWB ($e\approx0.9$),
make this a very unusual system. The thermal X-ray emission associated
with the wind collision region (WCR) and surrounding nebula has been
extremely well studied \citep[e.g.][and references
therein]{Hamaguchi2014} and considerable theoretical work has gone
into understanding this emission \citep[e.g.][]{EtaCar:Pittard02,
  Parkin2011, Madura2013}.

The LAT-detected emission above 200\,MeV has been reported by
\cite{EtaCar:Fermi10}, \cite{EtaCar:Farnier11}, and
\cite{Reitberger2012}.  The high-energy \g-ray spectrum exhibits two
distinct features: a low-energy component with a cutoff around 1\,GeV,
and a significantly harder component above $\sim$10\,GeV. Both
components are found to be variable, but on longer timescales and less
dramatically than in X-rays.  Upper limits from H.E.S.S.\ at energies
above $\sim$500\,GeV are more restrictive than the extrapolated
\fermi\ flux, implying a sudden drop in \g-ray flux
\citep{Abramowski2012}.  The high gas densities present in the WCR of
CWBs may result in the efficient production of $\pi^0$-decay \g\ rays
\citep{EtaCar:Farnier11, Bednarek2011}, and lead to a calorimetric
situation where {all} energy injected into particle acceleration is
radiated away on short timescales. In addition, the two shocks in this
system have very different properties, and cannot be considered as a
single system~\citep{Bednarek2011}.  The acceleration and interaction
of non-thermal particles in \etacar\ has been studied in previous work
\citep[e.g.][]{EtaCar:Agile, EtaCar:Fermi10, EtaCar:Farnier11,
  Bednarek2011}, but the complex geometry, and all of the relevant
phase-dependent timescales have so far not been fully considered. Here
we present a 3D dynamical model combined with particle injection,
propagation and interaction used to study the origin of \g-ray
emission from \etacar. We consider the stellar wind shocks of both
stars in \etacar, and model the light curve and spectra from MeV to
TeV energies. We first present the geometrical model and discuss the
relevant timescales in the system before moving to the full model,
results and discussion.

\section{Model}
\subsection{Dynamical model}

To model the non-thermal emission from \etacar\ we apply the dynamical
model introduced in \citet{Parkin2008}. The orbit of the two stars is
calculated in the centre-of-mass frame and the stellar winds are
assumed to have reached terminal velocities before they
interact. Orbital and stellar parameters are given in
Tab.~\ref{tab:stars}. Two regions are defined \citep{Parkin2008}:

\begin{table}
  \begin{center}
  \caption{Stellar parameters used throughout this work.}
  \label{tab:stars}
  \begin{tabular}{lccc}
    \hline
    Parameter & Primary & Companion & Reference \\\hline
    $R_* (R_\odot)$ & 100 & 20 & 1 \\
    $T_*$ (K) & 25800 & 30000 & 2 \\
    $L_* (10^6 L_\odot)$ & 4 & 0.3 & 2 \\
    $\dot{M} (M_\odot$\,yr$^{-1})$ & $4.8\times10^{-4}$ & $1.4\times10^{-5}$ & 3 \\
    $v_{\infty}$ (km\,s$^{-1}$) & 500 & 3000 & 3 \\\hline
  \end{tabular}
  \end{center}

  {\bf References:} (1) \citet{Hillier2001};
  (2) \citet{Davidson1997}; (3) \citet{EtaCar:Parkin09}.
\end{table}

{\bf The shock-cap} is the region where the two stellar winds collide,
and the flow accelerates outwards from the stagnation point along the
contact discontinuity (CD). The \textit{ballistic point} is defined
where the flow speed reaches 85\% of the primary terminal wind
speed. We assume radial symmetry w.r.t. to the apex, but the orbital
motion introduces a skew in the orientation of the shock-cap. The skew
angle changes as a function of orbital phase $\varphi$, and can reach
up to $38^\circ$ but is well below $10^\circ$ for
$0.1\lesssim \varphi \lesssim0.9$. The shape of the shock cap,
tangential velocity of the flow and surface density is determined by
momentum balance of the two winds \citep{Canto1996}. For 100 $\varphi$
bins from $\varphi=0.0$ to $\varphi=1.0$ we generate a grid of points
along the shock cap. This grid is formed by 36 azimuthal and 82 radial
points.

{\bf The ballistic flow} is entered beyond the ballistic point, where
the flow is assumed to be unaffected by the stars' gravity, ram
pressure of the winds or thermal pressure in the WCR. Beyond the
ballistic point, the wind material of the two stars is assumed to mix
over a mixing length (see below) and to flow with a speed according to
momentum balance of the two winds.

{\bf A collapse of the WCR} onto the companion star or dynamical
instabilities in the WCR were proposed to explain the dramatic
variability seen in thermal X-rays close to periastron
\citep[e.g.][]{Corcoran2005, Parkin2011}. The collapse is modelled by
turning off particle acceleration in the inner 80\% of the shock cap
during the X-ray minimum ($0.985 \lesssim \varphi \lesssim 1.025$).

\begin{figure}
  \includegraphics[width=0.95\columnwidth, trim=5 11 9 65,
  clip]{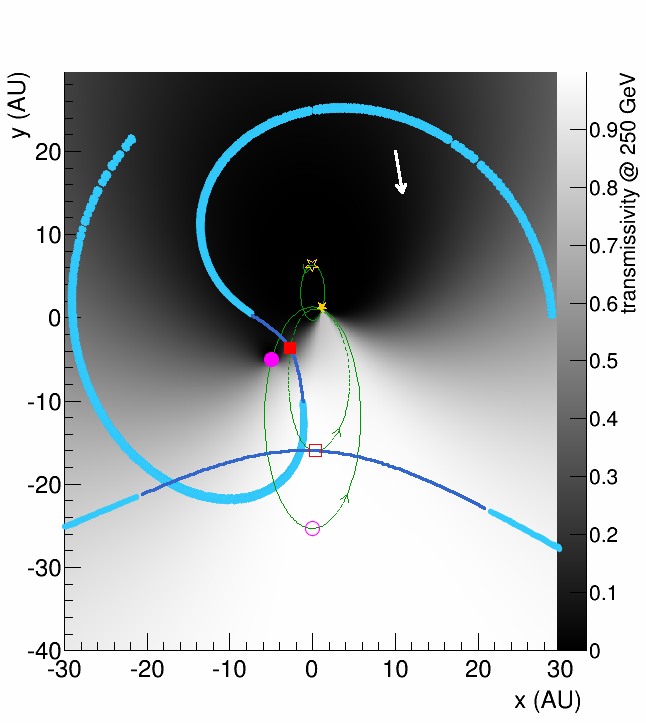}
  \caption{Geometry of the \etacar\ system as a function of orbital
    phase overlaid on a transmissivity map at 250\,GeV at
    $\varphi=0.04$ in the plane of the two stars. The green solid,
    long-dashed and short-lines show the motions of the primary,
    companion and stagnation point around the centre-of-mass of the
    system. At $\varphi=0.04$ and $\varphi=0.5$ the positions of these
    elements in the system are represented respectively by closed and
    open symbols. The position of the shock cap is indicated by a blue
    line, the ballistic part is shown as azure segments that show the
    direction of movement of each of the computational elements. The
    projected line of sight to the observer is shown with a thick
    black arrow \citep{Madura2012}.}
  \label{fig:geometry}
\end{figure}

Fig.~\ref{fig:geometry} shows the overall geometry of the system, the
motion of the two stars and the stagnation point of the two winds, the
shock cap size, geometry at two phases, and the line-of-sight to the
observer (projected onto the orbital plane). The asymmetric path of
the stagnation point is due to the skew of the shock cap caused by the
rapid motion of the companion around periastron.

\subsection{Timescales}

Several different timescales in \etacar\ are relevant for particle
acceleration and interaction. The acceleration time $t_\mathrm{acc}$
of a particle in DSA is determined by the shock speed $v_s$, the
magnetic field strength $B$, and the diffusion in terms of the Bohm
diffusion coefficient $\kappa=\eta_\mathrm{acc}\kappa_{B}$:
$ t_{\rm acc} \approx 50\,\eta_\mathrm{acc}\,v_{s,10^3\,{\rm
    km\,s}^{-1}}^{-2}(E_{\rm acc,GeV} / B_{\rm G})$\,s.

{\bf The magnetic field} of the two stars in \etacar\ is unknown. We
adopt plausible surface magnetic field strengths of 100\,G \citep[see
e.g.][]{Walder2012}, and spatial dependence following
\citet[][]{Eichler1993}, with a toroidal form for most of the orbit
and varying between $(0.1\lesssim B \lesssim 10)$\,G.

{\bf The post-shock (PS) regions} on both sides of the CD are very
different in nature: along the entire orbit the primary shock is
expected to be radiative, whereas the companion shock is
adiabatic. The slow dense primary wind collapses into a thin sheet of
dense material, while the gas on the companion side flows out of the
system without significant cooling. We calculate the thickness of the
primary PS region for a radiative shock following
\citet{Zhekov2007}. For the companion side, the thickness of the PS
region can be calculated from mass conservation. For the wind
parameters in Tab.~\ref{tab:stars}, the typical thickness of the
companion PS region ($\sim$1 AU) is orders of magnitude larger than
the primary PS region ($\sim$$10^{-4}$\,AU).
Note that densities in the WCR are typically
$10^8$\,cm$^{-3}$($10^7$\,cm$^{-3}$)
on the primary (companion) side at apastron, and an order of magnitude
higher close to periastron.

\begin{figure}
  \includegraphics[width=0.95\columnwidth, trim=5 0 50 30,
  clip]{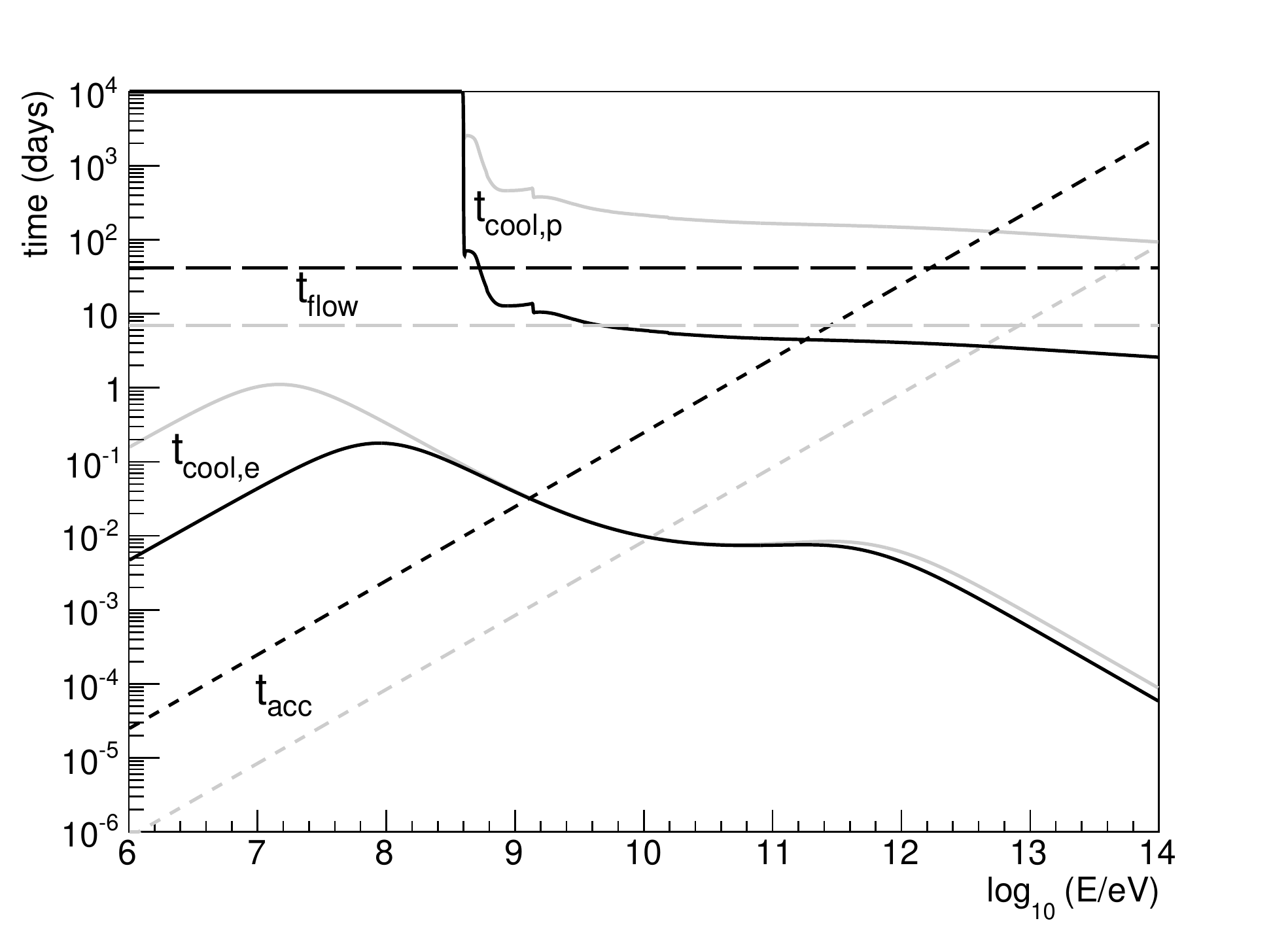}
  \caption{Key timescales in the primary (black) and companion (gray)
    shocked wind. The flow time of a gas packet to reach the shock cap
    edge is shown in long-dashed, the acceleration timescale in
    short-dashed, and the cooling timescales for electrons
    ($t_\mathrm{cool,e}$,
    Coulomb, IC, and synchrotron losses) and protons
    ($t_\mathrm{cool,p}$)
    in solid lines. $\eta_mathrm{acc}=15$ is assumed in both cases.}
    \label{fig:timescales}
\end{figure}

Fig.~\ref{fig:timescales} shows the key timescales as a function of
particle energy in the primary and companion PS regions for
$\varphi=0.05$. On both sides of the CD, electrons cool much faster
than the typical time it takes gas to flow from the shock apex to the
edge.Coulomb scattering dominates below 100\,MeV, while inverse
Compton (IC) scattering ($E_e\sim10$\,GeV) and synchrotron cooling
($E_e\gtrsim100$\,GeV) dominate at higher electron energies.

On the primary side the cooling time for protons is still shorter than
the flow time. This suggests that for this shock an equilibrium is
reached and that essentially no accelerated particles leave the shock
cap. The cooling time of protons on the companion side, however, is
longer than the flow time, which implies that protons are accelerated
whilst moving along the shock cap and escape in to the ballistic
flow. This more complex situation is treated separately in a
semi-analytic model described below.

\subsection{Time-dependent model}\label{sec:model}

The dynamical model implies that the solid angles of the shock-cap as
viewed from the primary and companion stars are 1.2 and 5.2\,sr,
respectively. In addition, only the wind kinetic power normal to the
shock will be available for particle acceleration, limiting the
available power to $1.6 (7.6) \times 10^{36}$\,\lumi\ on the primary
(companion) side. We model the two components of emission detected by
\fermi\ as arising from the two sides of the WCR as initially
suggested by \citet{Bednarek2011}. The low energy component, which is
extremely luminous and has a cutoff around a few GeV, originates on
the primary side, where the high PS density will limit particle
acceleration and provide high emission efficiency through the
interaction of all accelerated protons.  The harder, fainter
high-energy component originates on the companion side, where the
lower density will allow for acceleration limited only by the flow
timescale, and result in lower luminosity due to most particles
escaping without interacting. However, a certain level of mixing
between the two layers of the WCR is needed to reach the detected
emission levels, a phenomenon seen in hydrodynamical simulations of
\etacar\ \citep{Parkin2011}.

The faster cooling timescales of electrons
(Fig.~\ref{fig:timescales}), mean that much smaller values of
$\eta_\mathrm{acc}$ are required than for protons, with super-Bohm
acceleration needed for the high-energy component. Electron dominance
of either component requires an electron to proton ratio of more than
one (cf the commonly assumed 1\%) due to the additional electron
emission produced down to MeV energies. For these reasons we adopt the
hadronic scenario in the following.

Below we give a general description of how time-dependent particle
injection and \g-ray production is treated in our model. We employ the
radiative code used in \citet{Hinton2007a}.

{\bf Primary side:} Acceleration of particles in the PS region of the
primary is in saturation and counterbalanced by losses. The maximum
energy a particle can reach depends on the exact location at which it
enters the shock. Not only the magnetic field changes across the shock
cap, but also the shock velocity, as the angle between stellar wind
and shock cap is changing. At the same time, radiation energy
densities and gas density change. To calculate the emission from the
primary side of the shock cap, we calculate the tangential velocity
$v_t$ of the radiative layer for each annulus \citep{Canto1996}. The
shock velocity is given by
$v_s = \frac{3}{4}\sqrt{v_{\rm prim}^2 - v_t^2}$.  Inserting $v_s$
into $t_{\rm acc}$ and solving for the energy where
$t_\mathrm{acc}=t_\mathrm{cool}$ yields the maximum particle energy in
each annulus. The power available for particle acceleration
$P_{\rm avail,i} = \dot{E}_{\rm avail,i}$ depends on the solid angle
of the annulus,
$\Omega_i = 2\pi(\cos{\theta_{i-1}} - \cos{\theta_i})$, the primary
mass-loss rate and wind speed:
$P_{\rm avail,i} = \frac{1}{8\pi}\Omega_i \dot{M_1} v_{w}^2$. A
constant fraction $\epsilon_p$ of the available wind power is assumed
to go into accelerated protons
$P_{\rm p,i} = \epsilon_p P_{\rm avail,i}$. From this we calculate the
CR proton injection spectrum and equilibrium \g-ray spectrum in each
annulus \citep{Ginzburg1964, Zabalza2011}. Given the high densities in
the primary wind and PS region, there is an additional contribution to
the high-energy \g-ray emission from charged pion decay and subsequent
secondary electron emission.

{\bf Companion side:} As can be seen in Fig.~\ref{fig:timescales}, the
lower PS density on the companion side results in accelerated protons
losing only a small fraction of their energy to p-p collisions in the
shock-cap. DSA therefore takes place under changing conditions as the
relativistic particle population flows outwards in the shock-cap. To
approximate this acceleration, we follow the CR population through
each annulus outwards on the shock cap, and apply a semi-analytic
acceleration scheme as follows.

We follow the standard picture of DSA in non-relativistic shocks
\citep{Bell1978}: as particles enter the shock they are scattered from
downstream to upstream by the turbulent wake and from upstream to
downstream by the Alfv\'en waves generated by the energetic particles
themselves attempting to escape upstream. After a time $\Delta t$,
particles with initial energy of $E_0$ will have a final energy
$\ln(E_{\Delta t}/E_0)=t/t_\mathrm{acc}$, where
$t_\mathrm{acc}= 3 \eta_\mathrm{acc} \kappa_\mathrm{B} v_s^{-2}
r(r+1)/(r-1)$,
$r$ is the shock compression ratio, and $v_s$ is the velocity of the
shock. For each crossing there is a probability $4u_2/c$, where $u_2$
is the downstream flow velocity, of the particles being advected
downstream, resulting in a mean probability of remaining in the shock
after a time $\Delta t$ of
$\ln(P_{\Delta t}) = -3/(r-1) \ln(E_t/E_0)$.  This process results in
a downstream particle distribution with a power-law index of
$p_\mathrm{CR}=-(r+2)/(r-1)$, which under strong shock conditions
($r=4$) results in the well-known $p_\mathrm{CR}=-2$ index.

We derive the relativistic particle distributions along the shock-cap
by, at each time step $\Delta t$, injecting a fraction
$\epsilon^0_\mathrm{c}$ of the wind kinetic power in particles with
energy $E_0 = 1$\,GeV. The gain in energy due to acceleration over the
time $\Delta t$ is applied as above, and a fraction $(1-P_{\Delta t})$
of the particles at each given energy are lost downstream.  On
subsequent steps, only the particles remaining in the shock will
continue to be accelerated, whereas particles lost downstream will be
advected with the annulus until the ballistic flow is reached (note
that very little energy is lost in the downstream region to p-p
collisions, as can be seen in Fig.~\ref{fig:timescales}). In addition
to the energy injected in particles at $E_0$, additional power is
required to provide the particle energy gain. For the properties of
the shock on the companion side of \etacar, we have found that a total
kinetic power fraction of
$\epsilon_\mathrm{c} \sim 15 \epsilon^0_\mathrm{c}$ is used for
particle acceleration.

Towards the edge of the shock-cap, the reduced wind ram pressure owing
to shock obliquity, and the increasing CR pressure from the particles
being accelerated may cause a modification of the structure of the
shock, leading to nonlinear effects in acceleration \citep[see,
e.g.,][for a review]{Malkov2001}.  To account for this effect we adopt
the semi-analytic approach of non-linear DSA by \cite{Berezhko1999},
resulting in a hardening at the highest energies of up to
$p_\mathrm{CR} \simeq -1.75$. A deeper study of the acceleration
process in CWB, including shock modification, will be discussed in a
forthcoming paper \citep{Zabalza2015}.

{\bf Emission from the ballistic flow:} Given the low gas density in
the companion PS region, most of the accelerated protons will not
interact, but leave the shock cap in the ballistic flow. Simulations
indicate that the wind material is subject to numerous instabilities
and that the radiative layer of gas mixes with the wind material of
the two stars \citep{Parkin2011}. We assume no mixing of the two
stellar winds in the shock-cap region, but full mixing over a certain
mixing length beyond the ballistic point. The two stellar winds and
the radiative layer are assumed to form a region of mixed material at
the edge of the shock cap at a distance $r_0$ from the apex of the
shock with thickness $d_0$ and density
$\rho_0 = (d_{\rm ps,1}\rho_1 + d_{\rm ps,2}\rho_2 + d_{\rm
  wall}\rho_{\rm wall}) / d_0$.
In our model mixing occurs exponentially, over a characteristic scale
equal to the shock cap radius. While moving away from the ballistic
point, the density of the mixed material decreases as $(r_0/r)^2$. As
the mixed material flows outwards the two stars orbit each other and a
spiral structure of dense rings will form
(cf. Fig.~\ref{fig:geometry}). The \g-ray emission from p-p
interactions in the ballistic flow is calculated in time steps much
shorter than the interaction timescale, with the emission spectrum
fixed to that found for p-p emission at the shock cap edge. We note
that the diffusion timescale out of the flow region is always much
longer than the flow timescale for the adopted value of $\eta$, the
modelled $B$-field, and the energy range considered.

{\bf Pair-production absorption} is significant for emission above
$\sim$100\,GeV given the strong stellar radiation fields.  We
calculated the pair production opacity \citep[e.g.][]{Dubus2006} along
the line of sight from each point on the shock cap and ballistic flow
in each phase bin, considering the system geometry for the given phase
and the density and direction of the primary and companion stellar
radiation fields (with parameters as shown in Tab.~\ref{tab:stars}).
A transmission map for the orbital plane is shown in
Fig.~\ref{fig:geometry}, showing how 250\,GeV emission from behind the
stars is completely suppressed.

\section{Results}

\begin{figure*}
  \includegraphics[width=\textwidth]{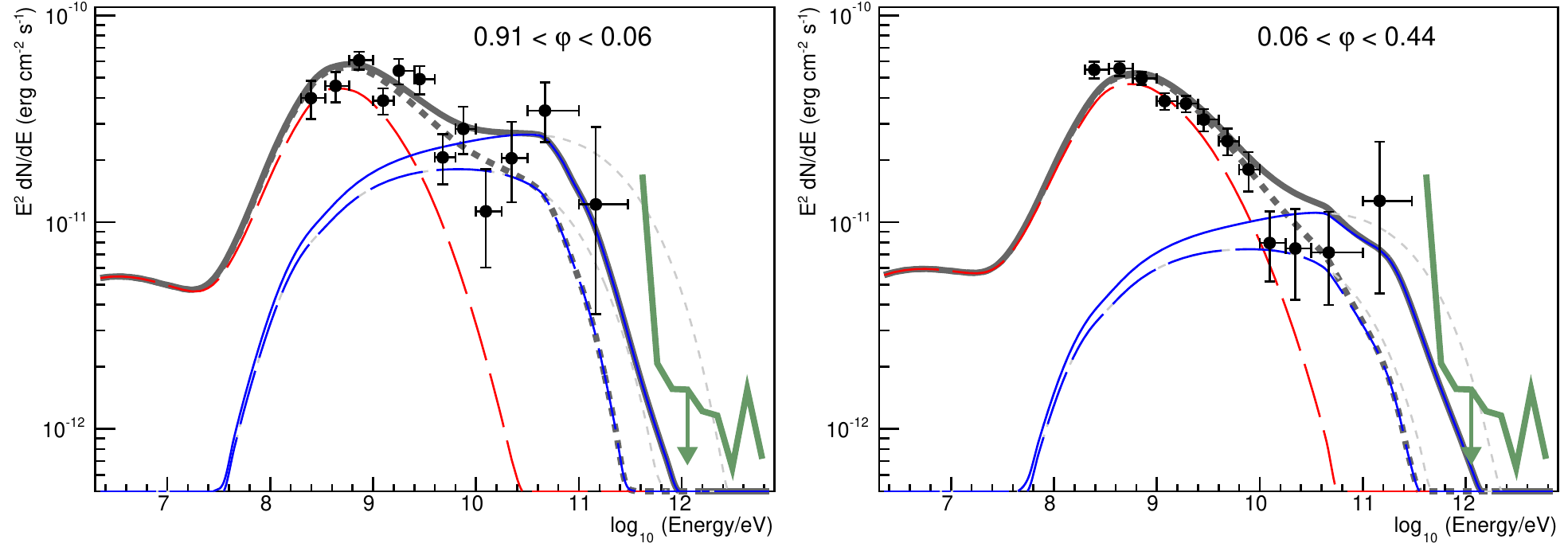} \caption{Model spectra
    and \g-ray SED for the periastron (left) and the apastron phase
    bin as used in \citet{Reitberger2012}. Red long-dashed lines
    indicate the primary contribution to the emission, assuming
    $\eta_\mathrm{acc}=15$. Solid (long-dashed) blue lines show the
    observed companion emission for $\eta_\mathrm{acc}=5$
    ($\eta_\mathrm{acc}=15$), and short-dashed thin gray lines show
    the intrinsic emission. Gray thick lines show the total emission
    for $\eta_\mathrm{acc}=5$ (solid) and $\eta_\mathrm{acc}=15$
    (short-dashed). TeV data are from \citet{Abramowski2012}.}
        \label{fig:sed}
\end{figure*}

Fig.~\ref{fig:sed} shows the \g-ray spectrum for two phase bins
including i) hadronic emission from accelerated protons on the primary
and companion side, ii) hadronic emission from protons accelerated in
the companion shock and interacting in the ballistic flow, and iii)
emission from secondary electrons on the primary side. A fraction of
$\epsilon_{\rm p,c} = 20\%$ of the available wind power goes in to
particle acceleration at both primary and companion shocks in the
curves shown and different values of $\eta$ are tested.

\begin{figure}
  \includegraphics[width=\columnwidth]{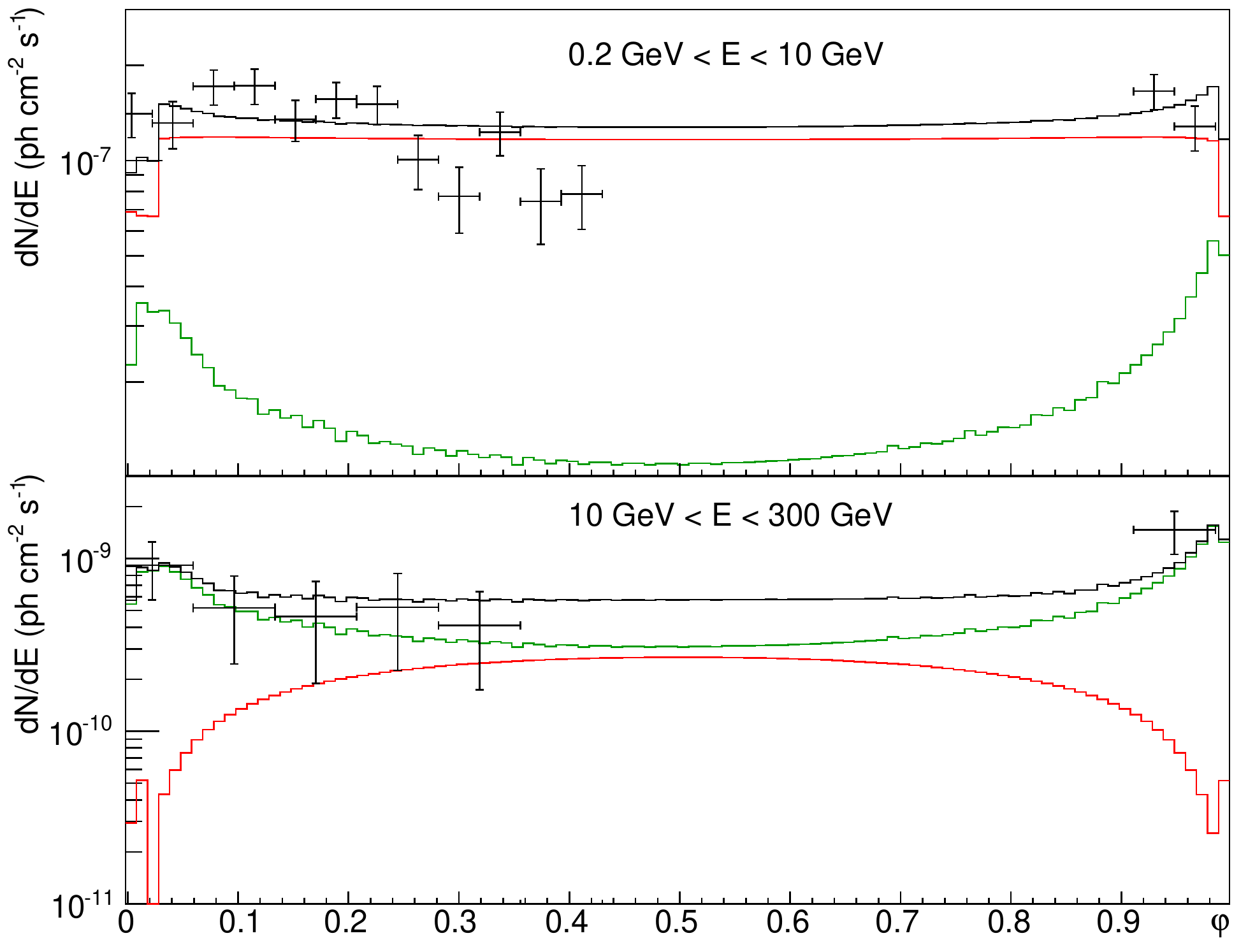}
  \caption{\g-ray lightcurve in the two energy bands used by
    \citet{Reitberger2012}. Red lines show the primary contribution,
    green the (absorbed) companion contribution, and black the total
    predicted emission. $\eta=15$ and $\epsilon=0.2$ are used. }
    \label{fig:lc}
\end{figure}
 
In the \g-ray lightcurve shown in Fig.~\ref{fig:lc} the low-energy
component is dominated by the primary at almost all
phases. Variability on the primary side is predicted only when the WCR
collapses. For the rest of the orbit, the \g-ray emission on the
primary side is constant due to the calorimetric behaviour and
constant injected power. \g-ray emission from the companion side is
variable over the orbit due to the changing densities in the ballistic
flow. The high-energy lightcurve is dominated by emission from protons
accelerated on the companion side that escape and interact in the
ballistic flow. Some residual \g-ray emission in both components of
the \g-ray spectrum are expected even during the collapse of the WCR
as a result of protons interacting in the ballistic flow that have
been launched at earlier phases at the ballistic point
(cf. Fig.~\ref{fig:geometry}).

\section{Discussion}

The model described above provides a reasonable level of agreement
with the observed \g-ray light curve and SED of \etacar. Furthermore,
the broad features of the expected emission emerge from simple
arguments based on the system geometry, mass flow and
energetics. Consideration of the time-dependent 3-D geometry of the
system is, however, critical for modelling of the light-curve around
periastron and the impact of pair-production absorption. We have shown
that inclusion of a realistic geometry is important for energetics and
the relative contribution of the two shocks and that the level of
emission around 1\,GeV requires extremely high-efficiency \g-ray
production, consistent (uniquely) with hadron calorimetry. We consider
the dominance of the SED by emission from accelerated protons and
nuclei to be robust. Dominance by electrons of either component is
very difficult due to energy-loss timescales and would certainly
require an electron to proton ratio of $>$1.

For the companion wind shock to reach the energies observed requires
either fairly efficient acceleration with $\eta_\mathrm{acc}\sim10$,
or much higher magnetic fields than the adopted stellar surface field
of 100\,G. To reach the required flux levels at $\sim$10\,GeV there
must be some mixing of the shocked and cooled primary wind with the PS
flow of the companion, resulting in $\mathcal{O}$(10\%) of accelerated
protons interacting. That mixing occurs on a characteristic scale of
the order of the shock cap size seems plausible \citep[see
e.g.][]{Parkin2011}.  The fact that acceleration in this shock
proceeds under changing conditions as particles flow around the shock
cap is intriguing, and analogous to the situation in very young
supernova remnants. Combined with the possibility of non-linear
effects and subsequent spectral hardening in this system, this fact
implies that \etacar\ may prove to be a very valuable system for
testing our ideas about particle acceleration. These acceleration
considerations are discussed in a forthcoming publication
\citep{Zabalza2015}.

The strong suppression of emission above $\sim$100\,GeV due to
pair-production absorption is unavoidable, given the constrained
geometry and stellar luminosities. However, the fact that in our
scenario the p-p emission is widely distributed over the mixing region
of the ballistic flow changes significantly the phase-dependence of
the absorption relative to the simplest
assumptions. \citet{Reitberger2012} introduced an external X-ray
absorber to explain the two-component \g-ray spectrum, but there is no
observational evidence for such an absorber. The energy density
required to cause a spectral feature in the SED is orders of magnitude
larger than the X-ray emission of any component in \etacar\
\citep{Hamaguchi2014}.

We are encouraged that with plausible assumptions and for the same
adopted parameters $\epsilon$ and $\eta$ for the two shocks, the model
provides reasonable agreement with the measured data. Our scenario can
be tested with better measurements around periastron, perhaps possible
with the lifetime \fermi\ data and with HESS-II and CTA. Agreement
with the measured light-curves is not possible without invoking some
kind of collapse around periastron, and is strongly motivated by the
X-ray observations.  The fact that in \etacar\ the highest energy
particles mostly escape from the system, and that this situation is
likely the same for other CWBs (with lower density winds) implies a
contribution of CWBs to the Galactic CRs up to TeV energies. However,
the fact that other known CWB systems typically have much lower
mass-loss rates also implies that such systems are likely very
difficult to detect in \g\ rays.

\bibliographystyle{mn2e_williams}
\bibliography{EtaCar}
\label{lastpage}

\end{document}